\documentclass[aps,pre,reprint,amsmath,amssymb,floats,longbibliography]{revtex4-2}
\usepackage{graphicx}
\usepackage{bm}
\usepackage{color}
\usepackage{physics}
\usepackage{longtable}
\usepackage{lipsum}

\begin{document}

\title{Operator growth in the transverse-field Ising spin chain with
integrability-breaking longitudinal field}

\author{Jae Dong Noh}
\affiliation{Department of Physics, University of Seoul, Seoul 02504,
Korea}

\date{\today}

\begin{abstract}
  We investigate the operator growth dynamics of the transverse field Ising
  spin chain in one dimension as varying the strength of the
  longitudinal field.  An operator in the Heisenberg picture spreads in the extended
  Hilbert space. Recently, it has been proposed that the spreading dynamics
  has a universal feature signaling chaoticity of underlying quantum
  dynamics. We demonstrate numerically that the operator growth dynamics in
  the presence of the longitudinal field follows the universal scaling law 
  for one-dimensional chaotic systems. We also find that the operator
  growth dynamics satisfies a crossover scaling law when the longitudinal
  field is weak. The crossover scaling confirms that the uniform longitudinal
  field makes the system chaotic at any nonzero value. We also discuss the
  implication of the crossover scaling on the thermalization
  dynamics and the effect of a nonuniform local longitudinal field.
\end{abstract}
\maketitle

\section{introduction}
There is growing interest in the theory for emergence of
equilibrium statistical mechanics in isolated quantum 
systems~\cite{D'Alessio.2016}. The canonical
typicality~\cite{Goldstein.2006}, a reincarnation of the quantum 
ergodic theory~\cite{Goldstein.2010, Reimann.2007}, assumes that a 
Hamiltonian eigenstate is statistically equivalent to a typical state in the
Hilbert space, so
a quantum mechanical expectation value of a local quantity is 
indistinguishable from the statistical ensemble average. The eigenstate
thermalization hypothesis~\cite{Srednicki.1999} makes an explicit and
testable ansatz for matrix elements of a local observable in the Hamiltonian
eigenstate basis to ensure the quantum thermalization.
Extensive numerical works have been performed to
examine the ansatz directly~(see Ref.~\cite{D'Alessio.2016} and 
references therein) and its thermodynamic implications on e.g. the
fluctuation-dissipation theorem~\cite{Khatami.2013,Noh.2020,Schuckert.2020,
Schonle.2021}.

Dynamical aspects of the quantum thermalization have also been 
attracting a growing interest. For example, out-of-time-ordered correlations
have been studied with the hope to uncover a chaotic signature of 
quantum 
dynamics~\cite{Maldacena.2016,Swingle.2018,Foini.2019,Murthy.2019,Brenes.2021}.
More recently, researchers gained insight into the quantum chaos from 
the operator growth dynamics. 
Quantum mechanics can be formulated in terms of the
time evolution of an operator in the Heisenberg picture. 
An operator, initially local and simple, becomes nonlocal and complex 
as it evolves in time, spreading in the operator Hilbert space. 
By quantifying and
characterizing the complexity of the operator growth dynamics, one may
have a better understanding of quantum chaos and equilibration dynamics 
of isolated quantum
systems~\cite{Gopalakrishnan.2018,Khemani.2018,Parker.2019,Dymarsky.2020,Susskind.2020,Avdoshkin.2020,Prosen.2007}.

The operator growth dynamics is intrinsically limited by an upper bound 
set by the spatial dimensionality and locality of interactions.
Parker {\em et al.} proposed a hypothesis that the operator growth
dynamics in nonintegrable systems follows a universal scaling law
corresponding to the maximal growth~\cite{Parker.2019}. The hypothesis is
supported by analytic and numerical calculations on the
Sachdev-Ye-Kitaev~(SYK)
model, which is defined in the infinite-dimensional space. Numerical results
on low-dimensional systems seem to be consistent with the hypothesis, but
more extensive studies are necessary for a decisive conclusion.

In this paper, we investigate numerically 
the operator growth dynamics in the transverse
field Ising~(t-Ising) spin chain in one dimension in the presence or absence
of a longitudinal field. The system is useful since one can control the
integrability by varying the longitudinal field~\cite{Kim.2014}.
We will show that the operator growth dynamics follows the universal scaling law predicted in
Ref.~\cite{Parker.2019}. The spatial structure of the one dimensional
lattice gives rise to a logarithmic correction in the operator growth
dynamics, which is absent in higher dimensional systems. 
Our results demonstrate the presence of the logarithmic correction. It
supports that the universal operator growth hypothesis~\cite{Parker.2019} 
is valid in low-dimensional systems.
We will also show that the operator growth dynamics
is an extremely useful tool for investigating the transition from 
integrability to nonintegrability. 
The system displays an interesting crossover as one turns on 
the uniform longitudinal field. Using the crossover, we will 
show that the system is thermal at any nonzero value of the longitudinal
field. As a byproduct, the crossover also reveals the scaling property 
of the thermalization
dynamics~\cite{Moeckel.2008,Gring.2012,Mallayya.2019,Mallayya.2018,Bertini.2015,Kollar.2011,Mori.2018}, which will be detailed in Sec.~\ref{sec4}.

The paper is organized as follows: In Sec.~\ref{sec2}, we present the
review on the operator growth dynamics. 
The universal feature of the operator
growth dynamics in some solvable models is summarized in 
Sec.~\ref{sec3}. 
We present our main results for the transverse field Ising spin chain in
Sec.~\ref{sec4}. Summary and discussions are given in Sec.~\ref{sec5}.

\section{Operator growth dynamics}\label{sec2}
We consider a system with Hamiltonian $H$ acting on the $D$ dimensional
Hilbert space. Focusing on operators instead of state vectors, one can
formulate the quantum mechanics with the von Neumann equation 
\begin{equation}
  \frac{\partial}{\partial t} O(t) = i \mathcal{L} O(t) \label{vNeq}
\end{equation}
for an operator $O(t)$ in the Heisenberg picture. Here, 
$\mathcal{L}$ is the Liouvillian superoperator defined as
\begin{equation}
  \mathcal{L} A = [H, A]  
  \label{Lop}
\end{equation}
with $\hbar=1$.
One can introduce an inner product between two operators. Then, an operator
$O(t)$ can be regarded as a state vector, denoted as $|O(t))$~\footnote{In
    order to distinguish the conventional state 
vector, we use the notation $|\cdot)$ instead of $\ket\cdot$ following
Ref.~\cite{Parker.2019}}, in the operator Hilbert space of dimensionality
$D^2$.
Under this point of view, the quantum mechanical dynamics describes the 
spreading or growth
of an initial state $|O(0))$ in the extended operator Hilbert space.

The operator growth is described conveniently with the Krylov basis.
An initial state $|O)$ spreads within the subspace spanned by 
$\{|O), \mathcal{L}|O), , \cdots\, \mathcal{L}^n |O),
\cdots\}$, called the Krylov subspace. The orthonormal basis set 
$\{|O_0), |O_1), \cdots, |O_n), \cdots \}$ can be constructed recursively 
using the Gram-Schmidt method. It starts with the normalized initial state
$|O_0) = |O)/ \sqrt{(O|O)}$ and proceeds to generate the successive 
basis states via the recursion relations
\begin{equation}
  \begin{aligned}
|A_n) &= \mathcal{L}|O_{n-1}) - b_{n-1} |O_{n-2}) \\
|O_n) &= \frac{1}{b_n} |A_n)  \mbox{ with } b_n = \sqrt{(A_n|A_n)} 
\end{aligned}
  \label{KB_recusrion}
\end{equation}
for $n\geq 1$. The operator inner product can be chosen as 
\begin{equation}
  (A|B) \equiv \frac{1}{D} {\rm Tr}\left[ A^\dagger B\right] .
  \label{inner_prod}
\end{equation}
It is the infinite temperature average of $A^\dagger B$. One may adopt a
different choice of the inner product~\cite{Parker.2019,Dymarsky.2020}.  
In this paper, however, we will take the simplest choice of
Eq.~\eqref{inner_prod}.

This procedure, which is usually referred to as the Lanczos
algorithm~\cite{Lanczos.1950}, results in the Krylov basis set and also the
sequence $\{b_n\}$, called the Lanczos coefficient with $b_0=0$.
In computational science, the Lanczos algorithm is one of the most important
numerical methods with which one can reduce a Hermitian matrix to a
tridiagonal form. It also underlies the recursion method which is 
a useful technique for evaluating the correlation functions in condensed matter 
physics.  For thorough reviews, we refer the
readers to Ref.~\cite{Viswanath.1994}.

Recently, Parker {\em et al.} attempted to use the Lanczos algorithm 
to characterize the operator growth dynamics~\cite{Parker.2019}.
An operator at time $t$ is written as
\begin{equation}
|O(t)) = e^{i \mathcal{L}t} |O_0) = \sum_{n=0}^\infty \varphi_n(t) |O_n) ,
    \label{formal_O} 
\end{equation}
where $\varphi_n(t) = (O_n|O(t))$ is the probability amplitude to be in the
$n$th Krylov state. The Liouvillian operator is represented as
a tridiagonal matrix $L_{m,n} = (O_m|\mathcal{L}|O_n)$ 
with $\mathcal{L}_{n, n-1} = \mathcal{L}_{n-1,n} = b_n$
and $\mathcal{L}_{n,m} = 0$ for $|n-m|\neq 1$. 
Thus, the probability amplitudes satisfy the discrete Schr\"odinger equation
\begin{equation}
    \dot\varphi_n(t) = b_n \varphi_{n-1} - b_{n+1} \varphi_{n+1}
    \label{Schrodinger_eq}
\end{equation}
with the initial condition $\varphi_n(0) = \delta_{n0}$ and $b_0=0$. 
Among all \{$\varphi_n(t)$\}, $\varphi_0(t)$ is equal to the autocorrelation
function $C_O(t) = (O|O(t))$.
The Schr\"odinger equation in Eq.~\eqref{Schrodinger_eq} describes 
a tight-binding system in a semi-infinite one-dimensional lattice 
with coordinate $n$, which will be called a depth in the Krylov space. 
Parker {\em et al.} showed that the Lanczos coefficient is bounded above 
for systems with local interactions in a $d$ dimensional space.
The bounds are
\begin{equation}
    b_n = \left\{
    \begin{aligned} &O(n / \ln n) \quad \mbox{ for } d = 1 \\
                    &O(n) \quad \mbox{ for } d>1 .
    \end{aligned}
\right.
    \label{maxbn}
\end{equation}
When the bound is achieved, the average depth $n_t \equiv \sum_n
n|\varphi_n(t)|^2$ grows fastest in time. That is, $n_t$ grows exponentially
in time when $b_n \propto n$, which signals the chaotic nature of quantum dynamics. 
Based on these observations and known results of solvable systems, they
hypothesize that the quantum systems are chaotic only when the Lanczos
coefficient follows the scaling law in
Eq.~\eqref{maxbn}~\cite{Parker.2019}.
There also exists a rigorous work on the lower bound for $\{b_n\}$ for a
specific class of systems including the chaotic Ising spin
chain~\cite{Cao.2021}.

\section{Solvable systems}\label{sec3}
\begin{table*}
  \caption{Lanczos coefficients and the autocorrelation function in exactly 
    solvable cases. $(\eta)_n= \eta(\eta+1)\cdots(\eta+n-1)$ is the Pochhammer
  symbol.}\label{table1}
  \begin{ruledtabular}
    \begin{tabular}{l|cccc}
      & $b_n$ & $C(t)$ & $\varphi_n(t)$ & $(n)_t = \sum_n
      n |\varphi_n(t)|^2 $ \\ \hline
      Type I & $\alpha$ & $J_1(2\alpha t)/(\alpha t) $ & 
      $ (n+1) J_{n+1}(2\alpha t)/(\alpha t)$ &
      $ \frac{16}{3\pi} \alpha t + o(t)$ \\
      Type II & $\alpha \sqrt{n}$ & $e^{-\alpha^2 t^2/2}$ & 
       $\frac{(\alpha t)^n}{\sqrt{n!}} e^{-\alpha^2 t^2/2}$ &
      $(\alpha t)^2$ \\
      Type III & $\alpha\sqrt{n(n-1+\eta)}$ & $
      (\sech \alpha t)^\eta$  & 
      $\sqrt{\frac{(\eta)n}{n!}} (\tanh \alpha t)^n (\sech \alpha
      t)^\eta$ & $\eta (\sinh \alpha t)^2 $ 
    \end{tabular}
  \end{ruledtabular}
\end{table*}

There are a few cases where the operator growth is exactly solvable.
We list the representative cases in Table~\ref{table1}. These cases are also
documented in Ref.~\cite{Viswanath.1994}, where the focus is put on the
analytic property of the autocorrelation function.

Consider first an artificial case with constant $b_n =
\alpha$~(type I). We are not aware of a local Hamiltonian and an observable
having the constant Lanczos coefficient. Nevertheless, it provides a useful
insight on the operator growth dynamics. 
We can rewrite the recursion relation in Eq.~\eqref{Schrodinger_eq} 
as $\varphi_{n-1} - \varphi_{n+1} = \frac{1}{\alpha} \dot\varphi_{n}$ 
for $n\geq 0$ requiring that $\varphi_{-1} \equiv 0$.
It has the same form as that of the Bessel functions, 
$J_{n-1}(x) - J_{n+1}(x) = 2J_{n}'(x)$~\cite{Arfken.2013}, 
except for the boundary term at $n=0$. The similarity suggests that
$\varphi_n(t)$ is the linear combination of the Bessel functions, $\varphi_n(t)
= \sum_{m\geq 0} c_m J_{n+m}(2\alpha t)$, whose coefficients are determined by
imposing that $\varphi_{-1}=0$. The resulting solution is $\varphi_n(t) =
J_n(2\alpha t) + J_{n+2}(2\alpha t) = (n+1)J_{n+1}(2\alpha t)/(\alpha t)$~(see
Table~\ref{table1}).
The correlation function $C(t) = \varphi_0(t) = 
J_1(2\alpha t)/(\alpha t)$ decays algebraically as 
$C(t) \simeq (\alpha t)^{-3/2} \cos(2\alpha t- 3\pi/4)$ in the long time
limit. 
It is straightforward to
evaluate the average depth $(n)_t = \sum_{n=0}^\infty n \varphi_n(t)^2$. It
grows linearly in time as $(n)_t = \frac{16}{3\pi}\alpha t + o(t)$. 

The spin-1/2 $XY$ chain exhibits similar behavior. The
autocorrelation function of the spin operator in the $z$ direction is given
by $C(t) = J_0(2\alpha t)^2 \simeq \frac{1}{\pi \alpha t} \cos(2\alpha t -
\pi/4)^2$~\cite{Niemeijer.1967,Cruz.1981} .
The Lanczos coefficient can be evaluated from the derivatives of $C(t)$ at
$t=0$~\cite{Viswanath.1994}. 
We evaluated numerically the Lanczos coefficient
and found that $b_n = \alpha + O(1/n)$, where the correction term
has an alternating sign. Since $b_n$ converges to a constant value, 
the average depth in the Krylov space scales linearly in time~(type I). 
The finite $n$ correction term determines the power-law decay exponent 
of $C(t)$ in the long time limit~\cite{Viswanath.1994}.

The second example~(type II) is realized when one considers the spin operator in
the $x$ direction in the spin-1/2 XY chain~\cite{Florencio.1987}. 
It also applies to the spin operator in the longitudinal direction in the
transverse field Ising spin chain. In this case, the depth $(n)_t = (\alpha
t)^2$ follows the quadratic scaling. It is faster than the linear
growth of type I, but still algebraic in time.

The last example~(type III) is characterized by the linear growth 
$b_n \sim \alpha n$ of the Lanczos coefficient and the exponential growth of
$(n)_t \sim e^{\alpha t}$. This case includes the SYK
model~\cite{Parker.2019,Roberts.2018} and the spin system on the 
two-dimensional lattice~\cite{Bouch.2015}. 
The exponent $\alpha$ is related to the positive Lyapunov exponent 
for the out-of-time-ordered
correlators~\cite{Parker.2019}. The latter is a signature of the quantum
chaos~\cite{Maldacena.2016}. Thus, the linear growth of $b_n$ 
can be regarded as a signature of quantum chaos. 

In one-dimensional systems with local interactions, the Lanczos coefficient
$b_n$ cannot grow linearly in $n$, but is constrained by the upper bound 
shown in Eq.~\eqref{maxbn}. There is no rigorous result confirming that the
upper bound is indeed achieved in a nonintegrable system. 
The specific
scaling with the logarithmic correction has not yet been confirmed 
numerically~\cite{Parker.2019}.
We will investigate the scaling behavior of the Lanczos
coefficient in the one-dimensional transverse field Ising spin chain
perturbed by the longitudinal field.

\section{Transverse and longitudinal field Ising spin chain}\label{sec4}
Consider lattice spins on an infinite one-dimensional lattice. Each spin at
site $l=0, \pm 1, \pm 2, \cdots$ is represented by the Pauli matrix 
$\sigma_l^a$ with $a = x,y$, and $z$.
Formally, the local
Pauli matrix $\sigma_l^a$ should be understood as the direct product
$\cdots \otimes I_{l-1} \otimes \sigma_l^a \otimes I_{l+1} \otimes \cdots$
with the identity operator $I_k$ at site $k$.
The Hamiltonian of the Ising model with
transverse and longitudinal fields~(tl-Ising model in short) is  given by
\begin{equation}
  H = J \sum_l \left[ \sigma_l^z \sigma_{l+1}^z + h \sigma_l^x + g_l
      \sigma_l^z\right] , 
  \label{H_tlIsing}
\end{equation}
where $J=1$ is the overall coupling constant, 
$h$ is a uniform transverse field, and $g_l$ is a site-dependent 
longitudinal field. 
The Ising model with only transverse field~(t-Ising model in short) is
equivalent to a free fermion system and integrable. 
The longitudinal field breaks the integrability and makes the system 
quantum chaotic~\cite{Kim.2014}. 

It is convenient to work with the basis set composed of the Pauli strings
of the form $\bm{\tau} \equiv \otimes_l \tau_l$ where $\tau_l \in
\{I_l, \sigma_l^x, \sigma_l^y, \sigma_l^z\}$ is a local operator acting on 
site $l$. 
The Pauli matrices have the property $\sigma^a \sigma^b = \delta_{ab} I + 
i \epsilon_{abc}\sigma^c$ with the Kronecker-$\delta$ symbol $\delta_{ab}$ 
and the  Levi-Civita symbol $\epsilon_{abc}$. This property guarantees that
the Pauli strings form the orthonormal set with
$({\bm \tau}|{\bm \tau'}) = \delta_{\bm{\tau}\bm{\tau'}}$.
A product of two Pauli strings is also a Pauli string with a possible phase factor.
Furthermore, for any pairs of Pauli strings $\bm{\tau}$ and $\bm{\tau'}$,
their products $\bm{\tau'}\bm{\tau}$ and $\bm{\tau}\bm{\tau'}$ are equal to
each other
up to a sign $(-1)^{\chi(\bm{\tau}, \bm{\tau' })}$, where $\chi$ counts the
number of sites where $\tau_l \neq I_l$, $\tau'_l \neq I_l$, and
$\tau_l\neq\tau'_l$. Consequently,
the commutator is given by $[{\bm \tau},{\bm \tau'}] = \left\{1-(-1)^{\chi({\bm
\tau},{\bm\tau'})}\right\} \bm{\tau \tau'}$. 
Using these algebraic properties of the Pauli strings, one can implement the
Lanczos algorithm easily. 
The operator algebra becomes even simpler by adopting the binary variable 
representation of a Pauli string. 
We refer the readers to Ref.~\cite{Dehaene.2003} 
and the Appendix of Ref.~\cite{Parker.2019} for more details.

As the initial operator $|O_0)$, we take a local one-body operator 
$O^a \equiv \sigma_0^a$ or a two-body operator $O^{aa}
\equiv \sigma_0^a \sigma_1^a$ with $a=x,y,z$~\footnote{One may
also consider $N$-body operators with $N>2$. We expect that the conclusion
would be the same as long as $N$ is finite.
}
When one applies the superoperator $\mathcal{L}$ to $|O_0)$ 
$n$ times, the spatial support of the resulting operator is of size 
$\xi = O(n)$. 
Thus, it is given by a linear superposition of
$O(4^\xi)$ Pauli strings. Due to the exponential increase,
a numerical computation of the Lanczos coefficient
is limited by the memory capacity of a computing system. 
In this paper, we report our results up to $n \le n_M$ with $n_M = 58$ for 
the t-Ising model and $38$ for the tl-Ising model.

\subsection{t-Ising model}
\begin{figure}[t]
    \includegraphics*[width=\columnwidth]{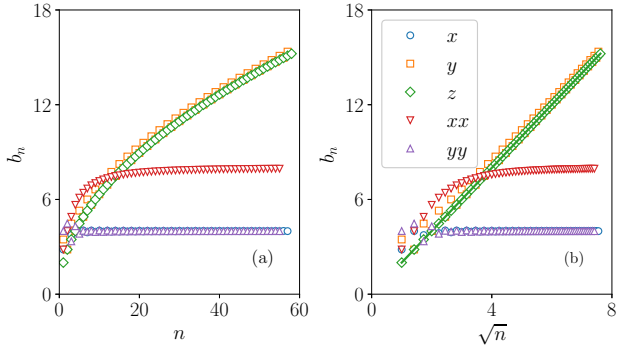}
    \caption{Plots of $b_n$ for the t-Ising model~($h=1$ and
        $g_l = 0$) with respect to $n$ in (a) and $\sqrt{n}$ in (b). $b_n$ of
      $O^z$ is in perfect agreement with the straight line in (b).}
    \label{fig1}
\end{figure}

We first present the results for the integrable t-Ising model 
with $h=1$ and $g_l=0$. Among six observables under consideration, 
$O^{y}$ and $O^{z}$ are characterized by the scaling law of type II. 
Figure~\ref{fig1}(b)
clearly demonstrates that $b_n \sim \sqrt{n}$ for $O^y$ and $O^z$.
For the other operators, the Lanczos coefficient converges to a constant
value. Note the t-Ising model with $h=1$ is
self-dual under the transformation $\tilde{\sigma}_l^x \leftrightarrow
\sigma_l^z \sigma_{l+1}^z$ and $\tilde{\sigma}_l^z \tilde{\sigma}_{l+1}^z
\leftrightarrow \sigma_{l+1}^x$. Thus, $O^x$ and $O^{zz}$ have the same
operator growth dynamics. We omit the plot of $O^{zz}$ in Fig.~\ref{fig1}.

The scaling behavior of the Lanczos coefficient is consistent with the
time dependence of the autocorrelation functions.
Brandt and Jacoby~\cite{Brandt.1976} derived  that 
$C_x(t) \equiv \langle \sigma_0^{x}(t)\sigma_0^x(t)\rangle = J_0(4t)^2 + 
J_1(4t)^2 \simeq \frac{1}{2\pi t} \left(1-\frac{\cos 8t}{8t}\right)$.
It decays algebraically with an oscillating component, which is a
characteristic of the operators of type I.
They also derived that $C_z(t) \equiv \langle 
\sigma_0^z(t)\sigma_0^{z}(0)\rangle 
= e^{-2 t^2}$, which shows that $O^z$ is an operator of type II with $\alpha =
2$. The Lanczos coefficient for $O^y$
also scales as $b_n \sim \sqrt{n}$ with an alternating finite-$n$ correction. 
The correction term indicates a power-law correction to the 
Gaussian autocorrelation function~\cite{Viswanath.1994}.

We also studied the t-Ising model with $h\neq 1$. 
We found that finite-$n$ corrections become larger, but the qualitative behavior
does not change. Summarizing the results, the Lanczos coefficients in the
integrable t-Ising model are of type I or II, depending on the choice of 
observables.

\subsection{tl-Ising model with uniform longitudinal field}
The longitudinal field breaks the integrability of the t-Ising
model~\cite{Kim.2014}. 
It is accepted that an integrable system becomes quantum chaotic immediately 
as a {\em uniform} integrability breaking field turns on. 
Various studies on energy-level spacing 
statistics~\cite{Rabson.2004,Santos.2010h1,Modak.2014,Modak.2014pxu} and
on the eigenstate thermalization hypothesis~\cite{Santos.2010h1,Rigol.2009}
confirm that a nonzero integrability breaking field results in quantum chaos.
Furthermore, the fidelity susceptibility measurement suggests that the
threshold value of an integrability breaking field necessary for the onset
of quantum chaos vanishes in the thermodynamic
limit~\cite{Pandey.2020,LeBlond.2020}. We will investigate the transition to
the quantum chaos by the uniform longitudinal field 
in the context of the operator growth.

It is conjectured that the operators of the one-dimensional chaotic systems
should follow the scaling law $b_n \propto n / W(n)$ with the Lambert $W$ function
$W(n) \simeq \ln n$~\cite{Parker.2019}. Numerical data 
for the tl-Ising model in Ref.~\cite{Parker.2019} seem to be consistent 
with the conjecture. However, the logarithmic corrections are not clearly visible 
in the data up to $n\lesssim 30$. 
In this subsection, we establish the scaling form $b_n \sim n / W(n)$ when the 
uniform longitudinal field $g_l = g$ is strong. We also investigate the 
crossover when $g$ is small.

\begin{figure}[t]
    \includegraphics*[width=\columnwidth]{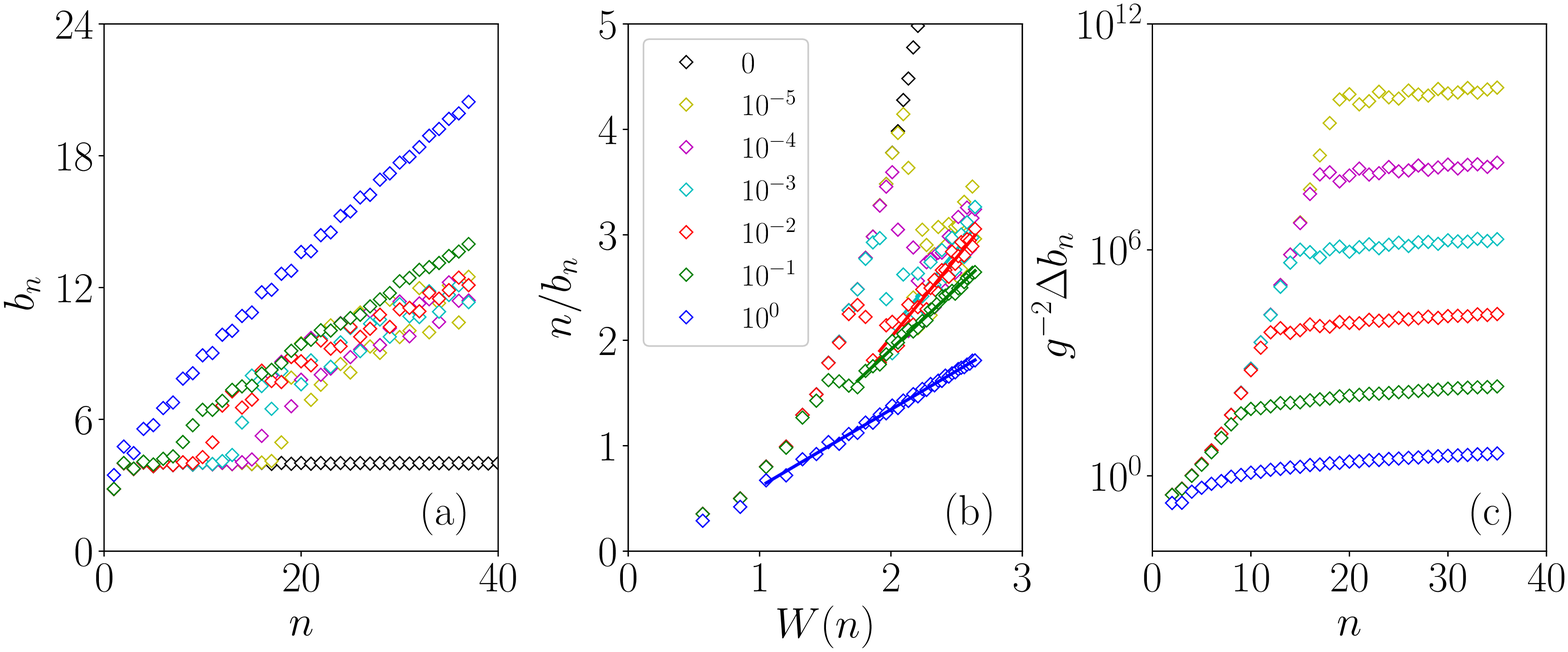}
    \caption{Lanczos coefficients for the operator $O^x=\sigma_0^x$ in 
      the tl-Ising model with $h=1$. (a) The  Lanczos coefficients at
      several values of the uniform longitudinal fields $g$ are compared. 
      (b) Plots of $n/b_n$ against $W(n)$. Data for $n>n_c(g)$ are in agreement
      with the straight lines, which indicates that $b_n \propto n/W(n)$ 
    for $n>n_c(g)$. (c) Scaling plot of $g^{-2} \Delta b_n$ vs $n$.}
    \label{fig2}
\end{figure}

We first report the results for the observable $O^x$ that exhibits the
scaling behavior of type I in the t-Ising model.
Numerical data are presented in Fig.~\ref{fig2}. 
The Lanczos coefficient increases with $n$ for $g\neq 0$. However, there is an 
overall downward curvature suggesting that the growth is sublinear.
It turns out that the logarithmic correction is responsible for the
curvature. In Fig.~\ref{fig2}(b), we plot $n/b_n$ as a function of $W(n)$. 
When $g=1$, the data are in excellent agreement with a straight line. 
It confirms the proposed scaling $b_n \propto n/W(n)$
for the one-dimensional quantum chaotic systems.

When the longitudinal field $g$ is weak, we find an interesting 
crossover at $n=n_c(g)$. The operator spreads as in 
the integrable system~($b_n(g) \simeq b_n(g=0)$) for small $n \ll n_c(g)$, 
then as in the chaotic system~($b_n \sim n/W(n)$) for $n \gg n_c(g)$. 
We have performed a quantitative analysis and found that 
\begin{equation}
  \Delta b_n \equiv \frac{b_n(g) - b_n(0)}{b_n(0)} \propto g^2 
    \label{h2_scaling}
\end{equation}
for $n \ll n_c(g)$.
Figure~\ref{fig2}(c) presents the plot of the scaled difference 
at several values of $g$. 
    The scaling plot demonstrates that the scaled differences $g^{-2}\Delta
    b_n$ from different values of $g$ lie on a single curve, represented
    by a scaling function $\mathcal{F}_x(n)$, until they cross over to the
    asymptotic behavior at $n \simeq n_c(g)$~[see Eq.~\eqref{scaling_form}].
Figure~\ref{fig2}(c) indicates 
that the scaling function has an exponential 
shape $\mathcal{F}_x(n) \sim e^{an}$ for large $n$ so that 
the crossover depth $n_c(g)$ scales as 
\begin{equation}
  n_c(g) \sim |\ln g| .
  \label{n_c} 
\end{equation}
The logarithmic dependence can be also inferred from the plots in
Figs.~\ref{fig2}(a) and \ref{fig2}(c), where the crossover points $n_c(g)$ 
are shifted by a constant amount per tenfold increase of $g$.

The crossover has an implication on the operator growth dynamics in the
Krylov space. At short times until $(n)_t$ reaches $n_c$, the operator 
spreads as in the integrable systems. 
Since the mean depth grows as $(n)_t \sim t$ in the type-I dynamics, 
the system reaches the crossover depth $n_c(g)$ at the crossover time 
$t_c$ 
\begin{equation}
    t_c(g) \sim n_c(g) \sim |\ln g| .
    \label{t_cross}
\end{equation}
Afterward, the generic spreading dynamics of the nonintegrable systems 
sets in.

The crossover explains the mechanism for the 
transition from prethermalization to thermalization.
When an integrable system is perturbed by an integrability breaking field, 
an observable temporarily remains at a nonthermal value predicted by 
the generalized Gibbs ensemble, and then tends to the thermal equilibrium 
value in the long time 
limit~\cite{Moeckel.2008,Gring.2012,Mallayya.2019,Mallayya.2018,Bertini.2015,Kollar.2011,Mori.2018}. 
The thermalization
dynamics is characterized by the 
rate which is proportional to the integrability breaking field
strength squared~\cite{Mallayya.2018,Mallayya.2019}. The thermalization
rate is manifest in the quadratic scaling in Eq.~\eqref{h2_scaling}. 
Besides the thermalization
rate, to the best of our knowledge, the crossover time following the scaling
law of Eq.~\eqref{t_cross} has not been reported yet.

\begin{figure}[t]
    \includegraphics*[width=\columnwidth]{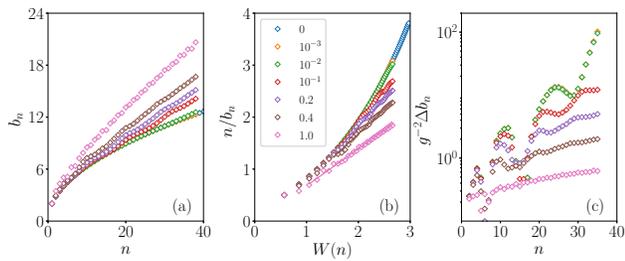}
    \caption{The same plots as in Fig.~\ref{fig2} for the operator 
    $O^z = \sigma_0^z$.}
    \label{fig3}
\end{figure}

We also report the results for the operator $O^z = \sigma_0^z$ in
Fig.~\ref{fig3}. The operator exhibits the scaling behavior of type II in the
integrable t-Ising model. Figures~\ref{fig3}(a) and \ref{fig3}(b) confirm that the 
Lanczos coefficient scales as $b_n \sim n / W(n)$ when the integrability
breaking field $g$ is large enough. The crossover also occurs for small $g$.
It is less trivial to locate the crossover depth $n_c(g)$ from the
numerical data in Figs.~\ref{fig3}(a) and \ref{fig3}(b). Nevertheless, we find that 
the scaling law in Eq.~\eqref{h2_scaling} is
also valid for the operator $O^z$, which is confirmed with the scaling plot in
Fig.~\ref{fig3}(c). The scaling implies that the thermalization 
rate is
also given by $\sim g^2$. 
Note that the scaling function for $O^z$ has a complicated shape with
oscillatory behavior, which makes it difficult to locate the crossover depth
$n_c(g)$.

\subsection{tl-Ising model with a longitudinal field at a single site}
Integrability can be broken with
a local perturbation~\cite{Santos.2004,Brenes.2020,Pandey.2020,Santos.2020}. 
For the integrable XXZ spin chain perturbed with a local magnetic field 
applied to a single site, the fidelity susceptibility measurement reveals
that the system becomes chaotic at any nonzero value of the magnetic field 
in the thermodynamic limit~\cite{Pandey.2020}. The t-Ising model has been
also studied with a local longitudinal field applied to
a single site~\cite{Santos.2020}.

In the perspective of the operator growth, it is surprising that the 
local perturbation leads to the quantum chaos. The operator growth in the
Krylov space is accompanied with the spatial growth of the operator support. 
With local perturbation, the support is affected minimally by a local
perturbation.
We investigate the impact of the local perturbation on the operator growth
dynamics in the tl-Ising model with the Hamiltonian in Eq.~\eqref{H_tlIsing}
with $h=1$ and $g_l = g \delta_{l0}$.

\begin{figure}[t]
    \includegraphics*[width=\columnwidth]{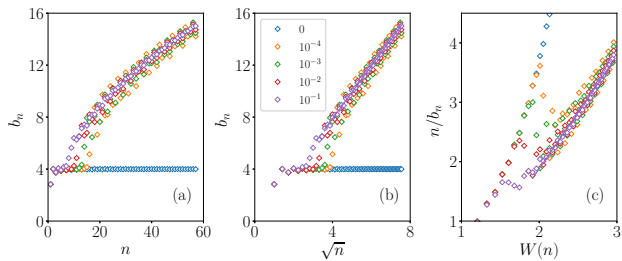}
    \caption{Lanczos coefficients for the operator $\sigma_0^x$ in 
        the tl-Ising model with $h=1$ and $g_l = g \delta_{l0}$ with $g\le
        0.1$. 
        (a) The  Lanczos coefficients at several values of the local 
        longitudinal fields $g$ are compared. There is a crossover from the
        type-I behavior.
        (b) Plots of $b_n$ against $\sqrt{n}$. The straight line represents 
        a linear fit of the data with $g=0.1$ for $n\geq 13$. 
    (c) Plots of $n/b_n$ against $W(n)$. The straight line also represents
    a linear fit of the data with $g=0.1$.}
    \label{fig4}
\end{figure}

Figure~\ref{fig4} presents the Lanczos coefficient for the operator $O^x =
\sigma_{0}^x$ when the local field strength $g \leq 10^{-1}$ is weak.
The operator $O^x$ follows the growth dynamics of type I without 
the longitudinal field. 
Figure~\ref{fig4}(a) looks similar to Fig.~\ref{fig2}(a).
The system undergoes a similar crossover at 
the depth $n= n_c(g) \propto |\ln g|$. On the other hand,
$b_n(g)$ for $n>n_c(g)$ shows a more pronounced downward curvature than in
Fig.~\ref{fig2}.
To characterize the asymptotic scaling behavior of $b_n$, 
we plot the Lanczos
coefficient with respect to $\sqrt{n}$ in Fig.~\ref{fig4}(b). The data for
large $n$ are well fitted to a straight line, which implies that 
$b_n \sim \sqrt{n}$, characteristic behavior of type-II dynamics. 
The asymptotic behavior, however, is not consistent with the quantum chaotic 
scaling $b_n \sim n/W(n)$. In
Fig.~\ref{fig4}(c), we present the plots of $n/b_n$ against the Lambert W
function $W(n)$.  The convex curvature invalidates the scaling form $b_n
\sim n / W(n)$. Thus, we conclude that the {\em weak} local perturbation 
is not sufficient to lead to the quantum chaos. It only modifies the
the operator growth dynamics from type I to type II.

\begin{figure}[t]
    \includegraphics*[width=\columnwidth]{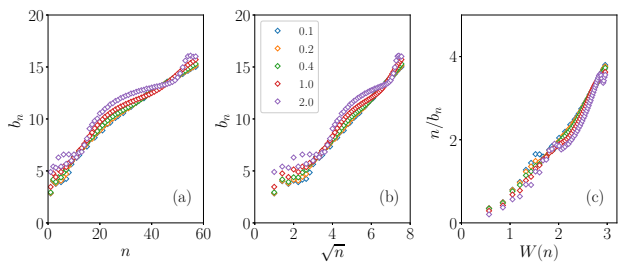}
    \caption{Lanczos coefficients for the operator $\sigma_0^x$ in 
        the tl-Ising model with $h=1$ and $g_l = g \delta_{l0}$ with
    $g\geq 0.1$. The data are plotted in the same way as in Fig.~\ref{fig4}.}
    \label{fig5}
\end{figure}

We also investigate the scaling behavior of $b_n$ when the local field 
strength is large. As $g$ increases, an oscillatory behavior sets in, which
obscures the asymptotic scaling behavior~(see Fig.~\ref{fig5}). 
The oscillatory behavior is reminiscent of the one observed in 
Fig.~\ref{fig3}. We speculate that the oscillatory behavior is a 
signature to a transition from the scaling of type II to the quantum chaotic
scaling. However, a decisive conclusion cannot be drawn from the numerical
data. 

We conclude that the weak local longitudinal field applied to
the t-Ising chain does not give rise to the quantum chaos: 
The threshold $g_c$ of the quantum chaos transition, if any, should be nonzero.
It is in contrast to the XXZ spin chain which undergoes an
immediate transition to the quantum chaos~\cite{Pandey.2020}.

\section{Summary and Discussions}\label{sec5}
We have investigated the operator growth dynamics in
the one-dimensional transverse-field Ising model perturbed with the uniform
and local longitudinal field. 
Without longitudinal field, the Lanczos coefficient $b_n$ 
converges to a constant~(type I) or scales as $O(\sqrt{n})$~(type II)
depending on the choice of local operators. When the longitudinal
field is uniform and strong enough, the Lanczos coefficient grows 
as $O(n/\ln n)$, which corresponds to the the maximum growth for a
one-dimensional system with local interactions. Our extensive numerical data in
Figs.~\ref{fig2} and \ref{fig3} confirm the existence of the logarithmic
correction to the linear scaling.
We were able to detect the logarithmic correction with the help
of the scaling analysis on the numerical data $b_n$ for large values of $n$.
These results support the hypothesis of Ref.~\cite{Parker.2019} that the
operator growth dynamics is a universal indicator of quantum chaos.

We have also discovered that the operator growth dynamics exhibits a crossover
scaling as the system undergoes a transition from an integrable
nonergodic state to a nonintegrable quantum chaotic state. When the uniform
longitudinal field strength $g$ is small, the Lanczos coefficients $b_n$ for
$n\ll n_c(g)$ follow the scaling form
\begin{equation}
    \frac{b_n(g)-b_n(0)}{b_n(0)} = g^2 \mathcal{F}_O(n)
    \label{scaling_form}
\end{equation}
with an operator-dependent scaling function $\mathcal{F}_O$~[see
Figs.~\ref{fig2}(c) and ~\ref{fig3}(c)]. For $n \gg n_c(g)$, $b_n$ crosses
over to the scaling form $O(n/\ln n)$. 
The crossover scaling is the irrefutable evidence that the integrability
breaking transition occurs $g_c = 0$.
The crossover scaling form is related
to the prethermalization dynamics. Since the Lanczos coefficient has
the dimension of the inverse time, the scaling factor $g^2$ corresponds to
the thermalization 
rate.
The crossover depth scales as $n_c(g) \sim |\ln
g|$ for the operator $\sigma_0^x$. The implication of the crossover depth on
the thermalization
dynamics has to be studied further.
The crossover scaling analysis also reveals that a local longitudinal field
at a single site does not give rise to the quantum chaos immediately. 

In conclusion, we establish that the operator growth dynamics faithfully 
reflects the quantum chaos in the transverse field Ising spin chain.
Furthermore, we show that it is a useful tool to characterize the transition of the
integrable system to the quantum chaos induced by the integrability 
breaking field.

\begin{acknowledgments}
  This work is supported by a National Research Foundation of Korea~(KRF)
  grant funded by the Korea government~(MSIP)~[Grant No. 2019R1A2C1009628].
\end{acknowledgments}

\bibliography{paper}

\begin{thebibliography}{52}%
\makeatletter
\providecommand \@ifxundefined [1]{%
 \@ifx{#1\undefined}
}%
\providecommand \@ifnum [1]{%
 \ifnum #1\expandafter \@firstoftwo
 \else \expandafter \@secondoftwo
 \fi
}%
\providecommand \@ifx [1]{%
 \ifx #1\expandafter \@firstoftwo
 \else \expandafter \@secondoftwo
 \fi
}%
\providecommand \natexlab [1]{#1}%
\providecommand \enquote  [1]{``#1''}%
\providecommand \bibnamefont  [1]{#1}%
\providecommand \bibfnamefont [1]{#1}%
\providecommand \citenamefont [1]{#1}%
\providecommand \href@noop [0]{\@secondoftwo}%
\providecommand \href [0]{\begingroup \@sanitize@url \@href}%
\providecommand \@href[1]{\@@startlink{#1}\@@href}%
\providecommand \@@href[1]{\endgroup#1\@@endlink}%
\providecommand \@sanitize@url [0]{\catcode `\\12\catcode `\$12\catcode
  `\&12\catcode `\#12\catcode `\^12\catcode `\_12\catcode `\%12\relax}%
\providecommand \@@startlink[1]{}%
\providecommand \@@endlink[0]{}%
\providecommand \url  [0]{\begingroup\@sanitize@url \@url }%
\providecommand \@url [1]{\endgroup\@href {#1}{\urlprefix }}%
\providecommand \urlprefix  [0]{URL }%
\providecommand \Eprint [0]{\href }%
\providecommand \doibase [0]{https://doi.org/}%
\providecommand \selectlanguage [0]{\@gobble}%
\providecommand \bibinfo  [0]{\@secondoftwo}%
\providecommand \bibfield  [0]{\@secondoftwo}%
\providecommand \translation [1]{[#1]}%
\providecommand \BibitemOpen [0]{}%
\providecommand \bibitemStop [0]{}%
\providecommand \bibitemNoStop [0]{.\EOS\space}%
\providecommand \EOS [0]{\spacefactor3000\relax}%
\providecommand \BibitemShut  [1]{\csname bibitem#1\endcsname}%
\let\auto@bib@innerbib\@empty
\bibitem [{\citenamefont {D'Alessio}\ \emph {et~al.}(2016)\citenamefont
  {D'Alessio}, \citenamefont {Kafri}, \citenamefont {Polkovnikov},\ and\
  \citenamefont {Rigol}}]{D'Alessio.2016}%
  \BibitemOpen
  \bibfield  {author} {\bibinfo {author} {\bibfnamefont {L.}~\bibnamefont
  {D'Alessio}}, \bibinfo {author} {\bibfnamefont {Y.}~\bibnamefont {Kafri}},
  \bibinfo {author} {\bibfnamefont {A.}~\bibnamefont {Polkovnikov}},\ and\
  \bibinfo {author} {\bibfnamefont {M.}~\bibnamefont {Rigol}},\ }\bibfield
  {title} {\bibinfo {title} {{From quantum chaos and eigenstate thermalization
  to statistical mechanics and thermodynamics}},\ }\href
  {https://doi.org/10.1080/00018732.2016.1198134} {\bibfield  {journal}
      {\bibinfo  {journal} {Adv. Phys.}\ }\textbf {\bibinfo {volume}
  {65}},\ \bibinfo {pages} {239} (\bibinfo {year} {2016})}\BibitemShut
  {NoStop}%
\bibitem [{\citenamefont {Goldstein}\ \emph {et~al.}(2006)\citenamefont
  {Goldstein}, \citenamefont {Lebowitz}, \citenamefont {Tumulka},\ and\
  \citenamefont {Zanghì}}]{Goldstein.2006}%
  \BibitemOpen
  \bibfield  {author} {\bibinfo {author} {\bibfnamefont {S.}~\bibnamefont
  {Goldstein}}, \bibinfo {author} {\bibfnamefont {J.~L.}\ \bibnamefont
  {Lebowitz}}, \bibinfo {author} {\bibfnamefont {R.}~\bibnamefont {Tumulka}},\
  and\ \bibinfo {author} {\bibfnamefont {N.}~\bibnamefont {Zanghì}},\
  }\bibfield  {title} {\bibinfo {title} {{Canonical Typicality}},\ }\href
  {https://doi.org/10.1103/physrevlett.96.050403} {\bibfield  {journal}
  {\bibinfo  {journal} {Phys. Rev. Lett.}\ }\textbf {\bibinfo {volume}
  {96}},\ \bibinfo {pages} {050403} (\bibinfo {year} {2006})}\BibitemShut
  {NoStop}%
\bibitem [{\citenamefont {Goldstein}\ \emph {et~al.}(2010)\citenamefont
  {Goldstein}, \citenamefont {Lebowitz}, \citenamefont {Tumulka},\ and\
  \citenamefont {Zanghì}}]{Goldstein.2010}%
  \BibitemOpen
  \bibfield  {author} {\bibinfo {author} {\bibfnamefont {S.}~\bibnamefont
  {Goldstein}}, \bibinfo {author} {\bibfnamefont {J.~L.}\ \bibnamefont
  {Lebowitz}}, \bibinfo {author} {\bibfnamefont {R.}~\bibnamefont {Tumulka}},\
  and\ \bibinfo {author} {\bibfnamefont {N.}~\bibnamefont {Zanghì}},\
  }\bibfield  {title} {\bibinfo {title} {{Long-time behavior of macroscopic
  quantum systems}},\ }\href {https://doi.org/10.1140/epjh/e2010-00007-7}
  {\bibfield  {journal} {\bibinfo  {journal} {Eur. Phys. J. H}\
  }\textbf {\bibinfo {volume} {35}},\ \bibinfo {pages} {173} (\bibinfo {year}
  {2010})}\BibitemShut {NoStop}%
\bibitem [{\citenamefont {Reimann}(2007)}]{Reimann.2007}%
  \BibitemOpen
  \bibfield  {author} {\bibinfo {author} {\bibfnamefont {P.}~\bibnamefont
  {Reimann}},\ }\bibfield  {title} {\bibinfo {title} {{Typicality for
  Generalized Microcanonical Ensembles}},\ }\href
  {https://doi.org/10.1103/physrevlett.99.160404} {\bibfield  {journal}
  {\bibinfo  {journal} {Phys. Rev. Lett.}\ }\textbf {\bibinfo {volume}
  {99}},\ \bibinfo {pages} {160404} (\bibinfo {year} {2007})}\BibitemShut
  {NoStop}%
\bibitem [{\citenamefont {Srednicki}(1999)}]{Srednicki.1999}%
  \BibitemOpen
  \bibfield  {author} {\bibinfo {author} {\bibfnamefont {M.}~\bibnamefont
  {Srednicki}},\ }\bibfield  {title} {\bibinfo {title} {{The approach to
  thermal equilibrium in quantized chaotic systems}},\ }\href
  {https://doi.org/10.1088/0305-4470/32/7/007} {\bibfield  {journal} {\bibinfo
      {journal} {J. Phys. A}\ }\textbf
  {\bibinfo {volume} {32}},\ \bibinfo {pages} {1163} (\bibinfo {year}
  {1999})}\BibitemShut {NoStop}%
\bibitem [{\citenamefont {Khatami}\ \emph {et~al.}(2013)\citenamefont
  {Khatami}, \citenamefont {Pupillo}, \citenamefont {Srednicki},\ and\
  \citenamefont {Rigol}}]{Khatami.2013}%
  \BibitemOpen
  \bibfield  {author} {\bibinfo {author} {\bibfnamefont {E.}~\bibnamefont
  {Khatami}}, \bibinfo {author} {\bibfnamefont {G.}~\bibnamefont {Pupillo}},
  \bibinfo {author} {\bibfnamefont {M.}~\bibnamefont {Srednicki}},\ and\
  \bibinfo {author} {\bibfnamefont {M.}~\bibnamefont {Rigol}},\ }\bibfield
  {title} {\bibinfo {title} {{Fluctuation-Dissipation Theorem in an Isolated
  System of Quantum Dipolar Bosons after a Quench}},\ }\href
  {https://doi.org/10.1103/physrevlett.111.050403} {\bibfield  {journal}
  {\bibinfo  {journal} {Phys. Rev. Lett.}\ }\textbf {\bibinfo {volume}
  {111}},\ \bibinfo {pages} {050403} (\bibinfo {year} {2013})}\BibitemShut
  {NoStop}%
\bibitem [{\citenamefont {Noh}\ \emph {et~al.}(2020)\citenamefont {Noh},
  \citenamefont {Sagawa},\ and\ \citenamefont {Yeo}}]{Noh.2020}%
  \BibitemOpen
  \bibfield  {author} {\bibinfo {author} {\bibfnamefont {J.~D.}\ \bibnamefont
  {Noh}}, \bibinfo {author} {\bibfnamefont {T.}~\bibnamefont {Sagawa}},\ and\
  \bibinfo {author} {\bibfnamefont {J.}~\bibnamefont {Yeo}},\ }\bibfield
  {title} {\bibinfo {title} {{Numerical Verification of the
  Fluctuation-Dissipation Theorem for Isolated Quantum Systems}},\ }\href
  {https://doi.org/10.1103/physrevlett.125.050603} {\bibfield  {journal}
  {\bibinfo  {journal} {Phys. Rev. Lett.}\ }\textbf {\bibinfo {volume}
  {125}},\ \bibinfo {pages} {050603} (\bibinfo {year} {2020})}\BibitemShut
  {NoStop}%
\bibitem [{\citenamefont {Schuckert}\ and\ \citenamefont
  {Knap}(2020)}]{Schuckert.2020}%
  \BibitemOpen
  \bibfield  {author} {\bibinfo {author} {\bibfnamefont {A.}~\bibnamefont
  {Schuckert}}\ and\ \bibinfo {author} {\bibfnamefont {M.}~\bibnamefont
  {Knap}},\ }\bibfield  {title} {\bibinfo {title} {{Probing eigenstate
  thermalization in quantum simulators via fluctuation-dissipation
  relations}},\ }\href {https://doi.org/10.1103/physrevresearch.2.043315}
  {\bibfield  {journal} {\bibinfo  {journal} {Phys. Rev. Res.}\
  }\textbf {\bibinfo {volume} {2}},\ \bibinfo {pages} {043315} (\bibinfo {year}
  {2020})}\BibitemShut {NoStop}%
\bibitem [{\citenamefont {Schönle}\ \emph {et~al.}(2021)\citenamefont
  {Schönle}, \citenamefont {Jansen}, \citenamefont {Heidrich-Meisner},\ and\
  \citenamefont {Vidmar}}]{Schonle.2021}%
  \BibitemOpen
  \bibfield  {author} {\bibinfo {author} {\bibfnamefont {C.}~\bibnamefont
  {Schönle}}, \bibinfo {author} {\bibfnamefont {D.}~\bibnamefont {Jansen}},
  \bibinfo {author} {\bibfnamefont {F.}~\bibnamefont {Heidrich-Meisner}},\ and\
  \bibinfo {author} {\bibfnamefont {L.}~\bibnamefont {Vidmar}},\ }\bibfield
  {title} {\bibinfo {title} {{Eigenstate thermalization hypothesis through the
  lens of autocorrelation functions}},\ }\href
  {https://doi.org/10.1103/physrevb.103.235137} {\bibfield  {journal} {\bibinfo
   {journal} {Phys. Rev. B}\ }\textbf {\bibinfo {volume} {103}},\ \bibinfo
  {pages} {235137} (\bibinfo {year} {2021})}\BibitemShut {NoStop}%
\bibitem [{\citenamefont {Maldacena}\ \emph {et~al.}(2016)\citenamefont
  {Maldacena}, \citenamefont {Shenker},\ and\ \citenamefont
  {Stanford}}]{Maldacena.2016}%
  \BibitemOpen
  \bibfield  {author} {\bibinfo {author} {\bibfnamefont {J.}~\bibnamefont
  {Maldacena}}, \bibinfo {author} {\bibfnamefont {S.~H.}\ \bibnamefont
  {Shenker}},\ and\ \bibinfo {author} {\bibfnamefont {D.}~\bibnamefont
  {Stanford}},\ }\bibfield  {title} {\bibinfo {title} {{A bound on chaos}},\
  }\href {https://doi.org/10.1007/jhep08(2016)106} {\bibfield  {journal}
      {\bibinfo  {journal} {J. High Energ. Phys.}\ }\textbf {\bibinfo
  {volume} {2016}},\ \bibinfo {pages} {106} (\bibinfo {year}
  {2016})}\BibitemShut {NoStop}%
\bibitem [{\citenamefont {Swingle}(2018)}]{Swingle.2018}%
  \BibitemOpen
  \bibfield  {author} {\bibinfo {author} {\bibfnamefont {B.}~\bibnamefont
  {Swingle}},\ }\bibfield  {title} {\bibinfo {title} {{Unscrambling the physics
  of out-of-time-order correlators}},\ }\href
  {https://doi.org/10.1038/s41567-018-0295-5} {\bibfield  {journal} {\bibinfo
  {journal} {Nat. Phys.}\ }\textbf {\bibinfo {volume} {14}},\ \bibinfo
  {pages} {988} (\bibinfo {year} {2018})}\BibitemShut {NoStop}%
\bibitem [{\citenamefont {Foini}\ and\ \citenamefont
  {Kurchan}(2019)}]{Foini.2019}%
  \BibitemOpen
  \bibfield  {author} {\bibinfo {author} {\bibfnamefont {L.}~\bibnamefont
  {Foini}}\ and\ \bibinfo {author} {\bibfnamefont {J.}~\bibnamefont
  {Kurchan}},\ }\bibfield  {title} {\bibinfo {title} {{Eigenstate
  thermalization hypothesis and out of time order correlators}},\ }\href
  {https://doi.org/10.1103/physreve.99.042139} {\bibfield  {journal} {\bibinfo
  {journal} {Phys. Rev. E}\ }\textbf {\bibinfo {volume} {99}},\ \bibinfo
  {pages} {042139} (\bibinfo {year} {2019})}\BibitemShut {NoStop}%
\bibitem [{\citenamefont {Murthy}\ and\ \citenamefont
  {Srednicki}(2019)}]{Murthy.2019}%
  \BibitemOpen
  \bibfield  {author} {\bibinfo {author} {\bibfnamefont {C.}~\bibnamefont
  {Murthy}}\ and\ \bibinfo {author} {\bibfnamefont {M.}~\bibnamefont
  {Srednicki}},\ }\bibfield  {title} {\bibinfo {title} {{Bounds on Chaos from
  the Eigenstate Thermalization Hypothesis}},\ }\href
  {https://doi.org/10.1103/physrevlett.123.230606} {\bibfield  {journal}
  {\bibinfo  {journal} {Phys. Rev. Lett.}\ }\textbf {\bibinfo {volume}
  {123}},\ \bibinfo {pages} {230606} (\bibinfo {year} {2019})}\BibitemShut
  {NoStop}%
\bibitem [{\citenamefont {Brenes}\ \emph {et~al.}(2021)\citenamefont {Brenes},
  \citenamefont {Pappalardi}, \citenamefont {Mitchison}, \citenamefont
  {Goold},\ and\ \citenamefont {Silva}}]{Brenes.2021}%
  \BibitemOpen
  \bibfield  {author} {\bibinfo {author} {\bibfnamefont {M.}~\bibnamefont
  {Brenes}}, \bibinfo {author} {\bibfnamefont {S.}~\bibnamefont {Pappalardi}},
  \bibinfo {author} {\bibfnamefont {M.~T.}\ \bibnamefont {Mitchison}}, \bibinfo
  {author} {\bibfnamefont {J.}~\bibnamefont {Goold}},\ and\ \bibinfo {author}
  {\bibfnamefont {A.}~\bibnamefont {Silva}},\ }\bibfield  {title} {\bibinfo
  {title} {{Out-of-time-order correlations and the fine structure of eigenstate
  thermalisation}},\ }\href@noop {} {\bibfield  {journal} {\bibinfo  {journal}
  {arXiv:2103.01161}\ } (\bibinfo {year} {2021})} \BibitemShut {NoStop}%
\bibitem [{\citenamefont {Gopalakrishnan}\ \emph {et~al.}(2018)\citenamefont
  {Gopalakrishnan}, \citenamefont {Huse}, \citenamefont {Khemani},\ and\
  \citenamefont {Vasseur}}]{Gopalakrishnan.2018}%
  \BibitemOpen
  \bibfield  {author} {\bibinfo {author} {\bibfnamefont {S.}~\bibnamefont
  {Gopalakrishnan}}, \bibinfo {author} {\bibfnamefont {D.~A.}\ \bibnamefont
  {Huse}}, \bibinfo {author} {\bibfnamefont {V.}~\bibnamefont {Khemani}},\ and\
  \bibinfo {author} {\bibfnamefont {R.}~\bibnamefont {Vasseur}},\ }\bibfield
  {title} {\bibinfo {title} {{Hydrodynamics of operator spreading and
  quasiparticle diffusion in interacting integrable systems}},\ }\href
  {https://doi.org/10.1103/physrevb.98.220303} {\bibfield  {journal} {\bibinfo
  {journal} {Phys. Rev. B}\ }\textbf {\bibinfo {volume} {98}},\ \bibinfo
  {pages} {220303} (\bibinfo {year} {2018})}\BibitemShut {NoStop}%
\bibitem [{\citenamefont {Khemani}\ \emph {et~al.}(2018)\citenamefont
  {Khemani}, \citenamefont {Vishwanath},\ and\ \citenamefont
  {Huse}}]{Khemani.2018}%
  \BibitemOpen
  \bibfield  {author} {\bibinfo {author} {\bibfnamefont {V.}~\bibnamefont
  {Khemani}}, \bibinfo {author} {\bibfnamefont {A.}~\bibnamefont
  {Vishwanath}},\ and\ \bibinfo {author} {\bibfnamefont {D.~A.}\ \bibnamefont
  {Huse}},\ }\bibfield  {title} {\bibinfo {title} {{Operator Spreading and the
  Emergence of Dissipative Hydrodynamics under Unitary Evolution with
  Conservation Laws}},\ }\href {https://doi.org/10.1103/physrevx.8.031057}
  {\bibfield  {journal} {\bibinfo  {journal} {Phys. Rev. X}\ }\textbf
  {\bibinfo {volume} {8}},\ \bibinfo {pages} {031057} (\bibinfo {year}
  {2018})}\BibitemShut {NoStop}%
\bibitem [{\citenamefont {Parker}\ \emph {et~al.}(2019)\citenamefont {Parker},
  \citenamefont {Cao}, \citenamefont {Avdoshkin}, \citenamefont {Scaffidi},\
  and\ \citenamefont {Altman}}]{Parker.2019}%
  \BibitemOpen
  \bibfield  {author} {\bibinfo {author} {\bibfnamefont {D.~E.}\ \bibnamefont
  {Parker}}, \bibinfo {author} {\bibfnamefont {X.}~\bibnamefont {Cao}},
  \bibinfo {author} {\bibfnamefont {A.}~\bibnamefont {Avdoshkin}}, \bibinfo
  {author} {\bibfnamefont {T.}~\bibnamefont {Scaffidi}},\ and\ \bibinfo
  {author} {\bibfnamefont {E.}~\bibnamefont {Altman}},\ }\bibfield  {title}
  {\bibinfo {title} {{A Universal Operator Growth Hypothesis}},\ }\href
  {https://doi.org/10.1103/physrevx.9.041017} {\bibfield  {journal} {\bibinfo
  {journal} {Phys. Rev. X}\ }\textbf {\bibinfo {volume} {9}},\ \bibinfo
  {pages} {041017} (\bibinfo {year} {2019})}\BibitemShut {NoStop}%
\bibitem [{\citenamefont {Dymarsky}\ and\ \citenamefont
  {Gorsky}(2020)}]{Dymarsky.2020}%
  \BibitemOpen
  \bibfield  {author} {\bibinfo {author} {\bibfnamefont {A.}~\bibnamefont
  {Dymarsky}}\ and\ \bibinfo {author} {\bibfnamefont {A.}~\bibnamefont
  {Gorsky}},\ }\bibfield  {title} {\bibinfo {title} {{Quantum chaos as
  delocalization in Krylov space}},\ }\href
  {https://doi.org/10.1103/physrevb.102.085137} {\bibfield  {journal} {\bibinfo
   {journal} {Phys. Rev. B}\ }\textbf {\bibinfo {volume} {102}},\ \bibinfo
  {pages} {085137} (\bibinfo {year} {2020})}\BibitemShut {NoStop}%
\bibitem [{\citenamefont {Susskind}(2020)}]{Susskind.2020}%
  \BibitemOpen
  \bibfield  {author} {\bibinfo {author} {\bibfnamefont {L.}~\bibnamefont
  {Susskind}},\ }\href {https://doi.org/10.1007/978-3-030-45109-7} {\emph
  {\bibinfo {title} {{Three Lectures on Complexity and Black Holes}}}},\
  SpringerBriefs in Physics\ (\bibinfo  {publisher} {Springer},\ 
      \bibinfo {address} {Cham}, \bibinfo
  {year} {2020})\BibitemShut {NoStop}%
\bibitem [{\citenamefont {Avdoshkin}\ and\ \citenamefont
  {Dymarsky}(2020)}]{Avdoshkin.2020}%
  \BibitemOpen
  \bibfield  {author} {\bibinfo {author} {\bibfnamefont {A.}~\bibnamefont
  {Avdoshkin}}\ and\ \bibinfo {author} {\bibfnamefont {A.}~\bibnamefont
  {Dymarsky}},\ }\bibfield  {title} {\bibinfo {title} {{Euclidean operator
  growth and quantum chaos}},\ }\href
  {https://doi.org/10.1103/physrevresearch.2.043234} {\bibfield  {journal}
  {\bibinfo  {journal} {Phys. Rev. Res.}\ }\textbf {\bibinfo {volume}
  {2}},\ \bibinfo {pages} {043234} (\bibinfo {year} {2020})}\BibitemShut
  {NoStop}%
\bibitem [{\citenamefont {Prosen}\ and\ \citenamefont
  {Žnidarič}(2007)}]{Prosen.2007}%
  \BibitemOpen
  \bibfield  {author} {\bibinfo {author} {\bibfnamefont {T.}~\bibnamefont
  {Prosen}}\ and\ \bibinfo {author} {\bibfnamefont {M.}~\bibnamefont
  {Žnidarič}},\ }\bibfield  {title} {\bibinfo {title} {{Is the efficiency of
  classical simulations of quantum dynamics related to integrability?}},\
  }\href {https://doi.org/10.1103/physreve.75.015202} {\bibfield  {journal}
  {\bibinfo  {journal} {Phys. Rev. E}\ }\textbf {\bibinfo {volume} {75}},\
  \bibinfo {pages} {015202(R)} (\bibinfo {year} {2007})}\BibitemShut {NoStop}%
\bibitem [{\citenamefont {Kim}\ \emph {et~al.}(2014)\citenamefont {Kim},
  \citenamefont {Ikeda},\ and\ \citenamefont {Huse}}]{Kim.2014}%
  \BibitemOpen
  \bibfield  {author} {\bibinfo {author} {\bibfnamefont {H.}~\bibnamefont
  {Kim}}, \bibinfo {author} {\bibfnamefont {T.~N.}\ \bibnamefont {Ikeda}},\
  and\ \bibinfo {author} {\bibfnamefont {D.~A.}\ \bibnamefont {Huse}},\
  }\bibfield  {title} {\bibinfo {title} {{Testing whether all eigenstates obey
  the eigenstate thermalization hypothesis}},\ }\href
  {https://doi.org/10.1103/physreve.90.052105} {\bibfield  {journal} {\bibinfo
  {journal} {Phys. Rev. E}\ }\textbf {\bibinfo {volume} {90}},\ \bibinfo
  {pages} {052105} (\bibinfo {year} {2014})}\BibitemShut {NoStop}%
\bibitem [{\citenamefont {Moeckel}\ and\ \citenamefont
  {Kehrein}(2008)}]{Moeckel.2008}%
  \BibitemOpen
  \bibfield  {author} {\bibinfo {author} {\bibfnamefont {M.}~\bibnamefont
  {Moeckel}}\ and\ \bibinfo {author} {\bibfnamefont {S.}~\bibnamefont
  {Kehrein}},\ }\bibfield  {title} {\bibinfo {title} {{Interaction Quench in
  the Hubbard Model}},\ }\href {https://doi.org/10.1103/physrevlett.100.175702}
  {\bibfield  {journal} {\bibinfo  {journal} {Phys. Rev. Lett.}\
  }\textbf {\bibinfo {volume} {100}},\ \bibinfo {pages} {175702} (\bibinfo
  {year} {2008})}\BibitemShut {NoStop}%
\bibitem [{\citenamefont {Gring}\ \emph {et~al.}(2012)\citenamefont {Gring},
  \citenamefont {Kuhnert}, \citenamefont {Langen}, \citenamefont {Kitagawa},
  \citenamefont {Rauer}, \citenamefont {Schreitl}, \citenamefont {Mazets},
  \citenamefont {Smith}, \citenamefont {Demler},\ and\ \citenamefont
  {Schmiedmayer}}]{Gring.2012}%
  \BibitemOpen
  \bibfield  {author} {\bibinfo {author} {\bibfnamefont {M.}~\bibnamefont
  {Gring}}, \bibinfo {author} {\bibfnamefont {M.}~\bibnamefont {Kuhnert}},
  \bibinfo {author} {\bibfnamefont {T.}~\bibnamefont {Langen}}, \bibinfo
  {author} {\bibfnamefont {T.}~\bibnamefont {Kitagawa}}, \bibinfo {author}
  {\bibfnamefont {B.}~\bibnamefont {Rauer}}, \bibinfo {author} {\bibfnamefont
  {M.}~\bibnamefont {Schreitl}}, \bibinfo {author} {\bibfnamefont
  {I.}~\bibnamefont {Mazets}}, \bibinfo {author} {\bibfnamefont {D.~A.}\
  \bibnamefont {Smith}}, \bibinfo {author} {\bibfnamefont {E.}~\bibnamefont
  {Demler}},\ and\ \bibinfo {author} {\bibfnamefont {J.}~\bibnamefont
  {Schmiedmayer}},\ }\bibfield  {title} {\bibinfo {title} {{Relaxation and
  Prethermalization in an Isolated Quantum System}},\ }\href
  {https://doi.org/10.1126/science.1224953} {\bibfield  {journal} {\bibinfo
  {journal} {Science}\ }\textbf {\bibinfo {volume} {337}},\ \bibinfo {pages}
  {1318} (\bibinfo {year} {2012})}\BibitemShut {NoStop}%
\bibitem [{\citenamefont {Mallayya}\ \emph {et~al.}(2019)\citenamefont
  {Mallayya}, \citenamefont {Rigol},\ and\ \citenamefont
  {Roeck}}]{Mallayya.2019}%
  \BibitemOpen
  \bibfield  {author} {\bibinfo {author} {\bibfnamefont {K.}~\bibnamefont
  {Mallayya}}, \bibinfo {author} {\bibfnamefont {M.}~\bibnamefont {Rigol}},\
  and\ \bibinfo {author} {\bibfnamefont {W.~D.}\ \bibnamefont {Roeck}},\
  }\bibfield  {title} {\bibinfo {title} {{Prethermalization and Thermalization
  in Isolated Quantum Systems}},\ }\href
  {https://doi.org/10.1103/physrevx.9.021027} {\bibfield  {journal} {\bibinfo
  {journal} {Phys. Rev. X}\ }\textbf {\bibinfo {volume} {9}},\ \bibinfo
  {pages} {021027} (\bibinfo {year} {2019})} \BibitemShut {NoStop}%
\bibitem [{\citenamefont {Mallayya}\ and\ \citenamefont
  {Rigol}(2018)}]{Mallayya.2018}%
  \BibitemOpen
  \bibfield  {author} {\bibinfo {author} {\bibfnamefont {K.}~\bibnamefont
  {Mallayya}}\ and\ \bibinfo {author} {\bibfnamefont {M.}~\bibnamefont
  {Rigol}},\ }\bibfield  {title} {\bibinfo {title} {{Quantum Quenches and
  Relaxation Dynamics in the Thermodynamic Limit}},\ }\href
  {https://doi.org/10.1103/physrevlett.120.070603} {\bibfield  {journal}
  {\bibinfo  {journal} {Phys. Rev. Lett.}\ }\textbf {\bibinfo {volume}
  {120}},\ \bibinfo {pages} {070603} (\bibinfo {year} {2018})}\BibitemShut
  {NoStop}%
\bibitem [{\citenamefont {Bertini}\ \emph {et~al.}(2015)\citenamefont
  {Bertini}, \citenamefont {Essler}, \citenamefont {Groha},\ and\ \citenamefont
  {Robinson}}]{Bertini.2015}%
  \BibitemOpen
  \bibfield  {author} {\bibinfo {author} {\bibfnamefont {B.}~\bibnamefont
  {Bertini}}, \bibinfo {author} {\bibfnamefont {F.~H.~L.}\ \bibnamefont
  {Essler}}, \bibinfo {author} {\bibfnamefont {S.}~\bibnamefont {Groha}},\ and\
  \bibinfo {author} {\bibfnamefont {N.~J.}\ \bibnamefont {Robinson}},\
  }\bibfield  {title} {\bibinfo {title} {{Prethermalization and Thermalization
  in Models with Weak Integrability Breaking}},\ }\href
  {https://doi.org/10.1103/physrevlett.115.180601} {\bibfield  {journal}
  {\bibinfo  {journal} {Phys. Rev. Lett.}\ }\textbf {\bibinfo {volume}
  {115}},\ \bibinfo {pages} {180601} (\bibinfo {year} {2015})}\BibitemShut
  {NoStop}%
\bibitem [{\citenamefont {Kollar}\ \emph {et~al.}(2011)\citenamefont {Kollar},
  \citenamefont {Wolf},\ and\ \citenamefont {Eckstein}}]{Kollar.2011}%
  \BibitemOpen
  \bibfield  {author} {\bibinfo {author} {\bibfnamefont {M.}~\bibnamefont
  {Kollar}}, \bibinfo {author} {\bibfnamefont {F.~A.}\ \bibnamefont {Wolf}},\
  and\ \bibinfo {author} {\bibfnamefont {M.}~\bibnamefont {Eckstein}},\
  }\bibfield  {title} {\bibinfo {title} {{Generalized Gibbs ensemble prediction
  of prethermalization plateaus and their relation to nonthermal steady states
  in integrable systems}},\ }\href {https://doi.org/10.1103/physrevb.84.054304}
  {\bibfield  {journal} {\bibinfo  {journal} {Phys. Rev. B}\ }\textbf
  {\bibinfo {volume} {84}},\ \bibinfo {pages} {054304} (\bibinfo {year}
  {2011})}\BibitemShut {NoStop}%
\bibitem [{\citenamefont {Mori}\ \emph {et~al.}(2018)\citenamefont {Mori},
  \citenamefont {Ikeda}, \citenamefont {Kaminishi},\ and\ \citenamefont
  {Ueda}}]{Mori.2018}%
  \BibitemOpen
  \bibfield  {author} {\bibinfo {author} {\bibfnamefont {T.}~\bibnamefont
  {Mori}}, \bibinfo {author} {\bibfnamefont {T.~N.}\ \bibnamefont {Ikeda}},
  \bibinfo {author} {\bibfnamefont {E.}~\bibnamefont {Kaminishi}},\ and\
  \bibinfo {author} {\bibfnamefont {M.}~\bibnamefont {Ueda}},\ }\bibfield
  {title} {\bibinfo {title} {{Thermalization and prethermalization in isolated
  quantum systems: a theoretical overview}},\ }\href
  {https://doi.org/10.1088/1361-6455/aabcdf} {\bibfield  {journal} {\bibinfo
          {journal} {J. Phys. B}\
  }\textbf {\bibinfo {volume} {51}},\ \bibinfo {pages} {112001} (\bibinfo
  {year} {2018})}\BibitemShut {NoStop}%
\bibitem [{Note1()}]{Note1}%
  \BibitemOpen
  \bibinfo {note} {To distinguish the conventional state vector, we
  use the notation $|\cdot )$ instead of $\ket \cdot $ following Ref.~\cite
  {Parker.2019}}\BibitemShut {NoStop}%
\bibitem [{\citenamefont {Lanczos}(1950)}]{Lanczos.1950}%
  \BibitemOpen
  \bibfield  {author} {\bibinfo {author} {\bibfnamefont {C.}~\bibnamefont
  {Lanczos}},\ }\bibfield  {title} {\bibinfo {title} {{An iteration method for
  the solution of the eigenvalue problem of linear differential and integral
  operators}},\ }\href {https://doi.org/10.6028/jres.045.026} {\bibfield
      {journal} {\bibinfo  {journal} {J. Res. Nat. Bureau Stand.}\ }\textbf {\bibinfo {volume} {45}},\ \bibinfo {pages} {255}
  (\bibinfo {year} {1950})}\BibitemShut {NoStop}%
\bibitem [{\citenamefont {Viswanath}\ and\ \citenamefont
  {Müller}(1994)}]{Viswanath.1994}%
  \BibitemOpen
  \bibfield  {author} {\bibinfo {author} {\bibfnamefont {V.~S.}\ \bibnamefont
  {Viswanath}}\ and\ \bibinfo {author} {\bibfnamefont {G.}~\bibnamefont
  {Müller}},\ }\href {https://doi.org/10.1007/978-3-540-48651-0} {\emph
  {\bibinfo {title} {{The Recursion Method, Application to Many-Body
  Dynamics}}}},\ Lecture Notes in Physics\ (\bibinfo  {publisher}
  {Springer-Verlag},\ \bibinfo {address} {Berlin},\ \bibinfo {year}
  {1994})\BibitemShut {NoStop}%
\bibitem [{\citenamefont {Cao}(2021)}]{Cao.2021}%
  \BibitemOpen
  \bibfield  {author} {\bibinfo {author} {\bibfnamefont {X.}~\bibnamefont
  {Cao}},\ }\bibfield  {title} {\bibinfo {title} {{A statistical mechanism for
  operator growth}},\ }\href {https://doi.org/10.1088/1751-8121/abe77c}
  {\bibfield  {journal} {\bibinfo  {journal} {J. Phys. A}\ }\textbf {\bibinfo {volume} {54}},\ \bibinfo
  {pages} {144001} (\bibinfo {year} {2021})}\BibitemShut {NoStop}%
\bibitem [{\citenamefont {Arfken}(2013)}]{Arfken.2013}%
  \BibitemOpen
  \bibfield  {author} {\bibinfo {author} {\bibfnamefont {G.~B.}\ \bibnamefont
  {Arfken}},\ }\href@noop {} {\emph {\bibinfo {title} {{Mathematical methods
  for physicists}}}},\ Academic press\ (\bibinfo  {publisher} {Academic
  press},\ \bibinfo{address} {London},\ \bibinfo {year} {2013})\BibitemShut {NoStop}%
\bibitem [{\citenamefont {Niemeijer}(1967)}]{Niemeijer.1967}%
  \BibitemOpen
  \bibfield  {author} {\bibinfo {author} {\bibfnamefont {T.}~\bibnamefont
  {Niemeijer}},\ }\bibfield  {title} {\bibinfo {title} {{Some exact
  calculations on a chain of spins 1/2}},\ }\href
  {https://doi.org/10.1016/0031-8914(67)90235-2} {\bibfield  {journal}
  {\bibinfo  {journal} {Physica}\ }\textbf {\bibinfo {volume} {36}},\ \bibinfo
  {pages} {377} (\bibinfo {year} {1967})}\BibitemShut {NoStop}%
\bibitem [{\citenamefont {Cruz}\ and\ \citenamefont
  {Goncalves}(1981)}]{Cruz.1981}%
  \BibitemOpen
  \bibfield  {author} {\bibinfo {author} {\bibfnamefont {H.~B.}\ \bibnamefont
  {Cruz}}\ and\ \bibinfo {author} {\bibfnamefont {L.~L.}\ \bibnamefont
  {Goncalves}},\ }\bibfield  {title} {\bibinfo {title} {{Time-dependent
  correlations of the one-dimensional isotropic XY model}},\ }\href
  {https://doi.org/10.1088/0022-3719/14/20/016} {\bibfield  {journal} {\bibinfo
      {journal} {J. Phys. C}\ }\textbf {\bibinfo
  {volume} {14}},\ \bibinfo {pages} {2785} (\bibinfo {year}
  {1981})}\BibitemShut {NoStop}%
\bibitem [{\citenamefont {Florencio}\ and\ \citenamefont
  {Lee}(1987)}]{Florencio.1987}%
  \BibitemOpen
  \bibfield  {author} {\bibinfo {author} {\bibfnamefont {J.}~\bibnamefont
  {Florencio}}\ and\ \bibinfo {author} {\bibfnamefont {M.~H.}\ \bibnamefont
  {Lee}},\ }\bibfield  {title} {\bibinfo {title} {{Relaxation functions, memory
  functions, and random forces in the one-dimensional spin-1/2 XY and
  transverse Ising models}},\ }\href {https://doi.org/10.1103/physrevb.35.1835}
  {\bibfield  {journal} {\bibinfo  {journal} {Phys. Rev. B}\ }\textbf
  {\bibinfo {volume} {35}},\ \bibinfo {pages} {1835} (\bibinfo {year}
  {1987})}\BibitemShut {NoStop}%
\bibitem [{\citenamefont {Roberts}\ \emph {et~al.}(2018)\citenamefont
  {Roberts}, \citenamefont {Stanford},\ and\ \citenamefont
  {Streicher}}]{Roberts.2018}%
  \BibitemOpen
  \bibfield  {author} {\bibinfo {author} {\bibfnamefont {D.~A.}\ \bibnamefont
  {Roberts}}, \bibinfo {author} {\bibfnamefont {D.}~\bibnamefont {Stanford}},\
  and\ \bibinfo {author} {\bibfnamefont {A.}~\bibnamefont {Streicher}},\
  }\bibfield  {title} {\bibinfo {title} {{Operator growth in the SYK model}},\
  }\href {https://doi.org/10.1007/jhep06(2018)122} {\bibfield  {journal}
      {\bibinfo  {journal} {J. High Energ. Phys.}\ }\textbf {\bibinfo
  {volume} {2018}},\ \bibinfo {pages} {122} (\bibinfo {year}
  {2018})}\BibitemShut {NoStop}%
\bibitem [{\citenamefont {Bouch}(2015)}]{Bouch.2015}%
  \BibitemOpen
  \bibfield  {author} {\bibinfo {author} {\bibfnamefont {G.}~\bibnamefont
  {Bouch}},\ }\bibfield  {title} {\bibinfo {title} {{Complex-time singularity
  and locality estimates for quantum lattice systems}},\ }\href
  {https://doi.org/10.1063/1.4936209} {\bibfield  {journal} {\bibinfo
      {journal} {J. Math. Phys.}\ }\textbf {\bibinfo {volume}
  {56}},\ \bibinfo {pages} {123303} (\bibinfo {year} {2015})}\BibitemShut
  {NoStop}%
\bibitem [{\citenamefont {Dehaene}\ and\ \citenamefont
  {Moor}(2003)}]{Dehaene.2003}%
  \BibitemOpen
  \bibfield  {author} {\bibinfo {author} {\bibfnamefont {J.}~\bibnamefont
  {Dehaene}}\ and\ \bibinfo {author} {\bibfnamefont {B.~D.}\ \bibnamefont
  {Moor}},\ }\bibfield  {title} {\bibinfo {title} {{Clifford group, stabilizer
  states, and linear and quadratic operations over GF(2)}},\ }\href
  {https://doi.org/10.1103/physreva.68.042318} {\bibfield  {journal} {\bibinfo
  {journal} {Phys. Rev. A}\ }\textbf {\bibinfo {volume} {68}},\ \bibinfo
  {pages} {042318} (\bibinfo {year} {2003})}\BibitemShut {NoStop}%
\bibitem [{Note2()}]{Note2}%
  \BibitemOpen
  \bibinfo {note} {{\protect One may also consider $N$-body
  operators with $N>2$. We expect that the conclusion would be the same as long
  as $N$ is finite.}}\BibitemShut {Stop}%
\bibitem [{\citenamefont {Brandt}\ and\ \citenamefont
  {Jacoby}(1976)}]{Brandt.1976}%
  \BibitemOpen
  \bibfield  {author} {\bibinfo {author} {\bibfnamefont {U.}~\bibnamefont
  {Brandt}}\ and\ \bibinfo {author} {\bibfnamefont {K.}~\bibnamefont
  {Jacoby}},\ }\bibfield  {title} {\bibinfo {title} {{Exact results for the
  dynamics of one-dimensional spin-systems}},\ }\href
  {https://doi.org/10.1007/bf01320179} {\bibfield  {journal} {\bibinfo
      {journal} {Z. Phys. B}\ }\textbf {\bibinfo
  {volume} {25}},\ \bibinfo {pages} {181} (\bibinfo {year} {1976})}\BibitemShut
  {NoStop}%
\bibitem [{\citenamefont {Rabson}\ \emph {et~al.}(2004)\citenamefont {Rabson},
  \citenamefont {Narozhny},\ and\ \citenamefont {Millis}}]{Rabson.2004}%
  \BibitemOpen
  \bibfield  {author} {\bibinfo {author} {\bibfnamefont {D.~A.}\ \bibnamefont
  {Rabson}}, \bibinfo {author} {\bibfnamefont {B.~N.}\ \bibnamefont
  {Narozhny}},\ and\ \bibinfo {author} {\bibfnamefont {A.~J.}\ \bibnamefont
  {Millis}},\ }\bibfield  {title} {\bibinfo {title} {{Crossover from Poisson to
  Wigner-Dyson level statistics in spin chains with integrability breaking}},\
  }\href {https://doi.org/10.1103/physrevb.69.054403} {\bibfield  {journal}
  {\bibinfo  {journal} {Phys. Rev. B}\ }\textbf {\bibinfo {volume} {69}},\
  \bibinfo {pages} {054403} (\bibinfo {year} {2004})}\BibitemShut {NoStop}%
\bibitem [{\citenamefont {Santos}\ and\ \citenamefont
  {Rigol}(2010)}]{Santos.2010h1}%
  \BibitemOpen
  \bibfield  {author} {\bibinfo {author} {\bibfnamefont {L.~F.}\ \bibnamefont
  {Santos}}\ and\ \bibinfo {author} {\bibfnamefont {M.}~\bibnamefont {Rigol}},\
  }\bibfield  {title} {\bibinfo {title} {{Onset of quantum chaos in
  one-dimensional bosonic and fermionic systems and its relation to
  thermalization}},\ }\href {https://doi.org/10.1103/physreve.81.036206}
  {\bibfield  {journal} {\bibinfo  {journal} {Phys. Rev. E}\ }\textbf
  {\bibinfo {volume} {81}},\ \bibinfo {pages} {036206} (\bibinfo {year}
  {2010})}\BibitemShut {NoStop}%
\bibitem [{\citenamefont {Modak}\ \emph {et~al.}(2014)\citenamefont {Modak},
  \citenamefont {Mukerjee},\ and\ \citenamefont {Ramaswamy}}]{Modak.2014}%
  \BibitemOpen
  \bibfield  {author} {\bibinfo {author} {\bibfnamefont {R.}~\bibnamefont
  {Modak}}, \bibinfo {author} {\bibfnamefont {S.}~\bibnamefont {Mukerjee}},\
  and\ \bibinfo {author} {\bibfnamefont {S.}~\bibnamefont {Ramaswamy}},\
  }\bibfield  {title} {\bibinfo {title} {{Universal power law in crossover from
  integrability to quantum chaos}},\ }\href
  {https://doi.org/10.1103/physrevb.90.075152} {\bibfield  {journal} {\bibinfo
  {journal} {Phys. Rev. B}\ }\textbf {\bibinfo {volume} {90}},\ \bibinfo
  {pages} {075152} (\bibinfo {year} {2014})}\BibitemShut {NoStop}%
\bibitem [{\citenamefont {Modak}\ and\ \citenamefont
  {Mukerjee}(2014)}]{Modak.2014pxu}%
  \BibitemOpen
  \bibfield  {author} {\bibinfo {author} {\bibfnamefont {R.}~\bibnamefont
  {Modak}}\ and\ \bibinfo {author} {\bibfnamefont {S.}~\bibnamefont
  {Mukerjee}},\ }\bibfield  {title} {\bibinfo {title} {{Finite size scaling in
  crossover among different random matrix ensembles in microscopic lattice
  models}},\ }\href {https://doi.org/10.1088/1367-2630/16/9/093016} {\bibfield
      {journal} {\bibinfo  {journal} {New J. Phys.}\ }\textbf {\bibinfo
  {volume} {16}},\ \bibinfo {pages} {093016} (\bibinfo {year}
  {2014})}\BibitemShut {NoStop}%
\bibitem [{\citenamefont {Rigol}(2009)}]{Rigol.2009}%
  \BibitemOpen
  \bibfield  {author} {\bibinfo {author} {\bibfnamefont {M.}~\bibnamefont
  {Rigol}},\ }\bibfield  {title} {\bibinfo {title} {{Breakdown of
  Thermalization in Finite One-Dimensional Systems}},\ }\href
  {https://doi.org/10.1103/physrevlett.103.100403} {\bibfield  {journal}
  {\bibinfo  {journal} {Phys. Rev. Lett.}\ }\textbf {\bibinfo {volume}
  {103}},\ \bibinfo {pages} {100403} (\bibinfo {year} {2009})}\BibitemShut
  {NoStop}%
\bibitem [{\citenamefont {Pandey}\ \emph {et~al.}(0 10)\citenamefont {Pandey},
  \citenamefont {Claeys}, \citenamefont {Campbell}, \citenamefont
  {Polkovnikov},\ and\ \citenamefont {Sels}}]{Pandey.2020}%
  \BibitemOpen
  \bibfield  {author} {\bibinfo {author} {\bibfnamefont {M.}~\bibnamefont
  {Pandey}}, \bibinfo {author} {\bibfnamefont {P.~W.}\ \bibnamefont {Claeys}},
  \bibinfo {author} {\bibfnamefont {D.~K.}\ \bibnamefont {Campbell}}, \bibinfo
  {author} {\bibfnamefont {A.}~\bibnamefont {Polkovnikov}},\ and\ \bibinfo
  {author} {\bibfnamefont {D.}~\bibnamefont {Sels}},\ }\bibfield  {title}
  {\bibinfo {title} {{Adiabatic Eigenstate Deformations as a Sensitive Probe
  for Quantum Chaos}},\ }\href {https://doi.org/10.1103/physrevx.10.041017}
  {\bibfield  {journal} {\bibinfo  {journal} {Phys. Rev. X}\ }\textbf
  {\bibinfo {volume} {10}},\ \bibinfo {pages} {041017} (\bibinfo {year}
  {2020})}\BibitemShut {NoStop}%
\bibitem [{\citenamefont {LeBlond}\ \emph {et~al.}(2020)\citenamefont
  {LeBlond}, \citenamefont {Sels}, \citenamefont {Polkovnikov},\ and\
  \citenamefont {Rigol}}]{LeBlond.2020}%
  \BibitemOpen
  \bibfield  {author} {\bibinfo {author} {\bibfnamefont {T.}~\bibnamefont
  {LeBlond}}, \bibinfo {author} {\bibfnamefont {D.}~\bibnamefont {Sels}},
  \bibinfo {author} {\bibfnamefont {A.}~\bibnamefont {Polkovnikov}},\ and\
  \bibinfo {author} {\bibfnamefont {M.}~\bibnamefont {Rigol}},\ }\bibfield
  {title} {\bibinfo {title} {{Universality in the Onset of Quantum Chaos in
  Many-Body Systems}},\ }\href@noop {} {\bibfield  {journal} {\bibinfo
  {journal} {arXiv:2012.07849}\ } (\bibinfo {year} {2020})} \BibitemShut {NoStop}%
\bibitem [{\citenamefont {Santos}(2004)}]{Santos.2004}%
  \BibitemOpen
  \bibfield  {author} {\bibinfo {author} {\bibfnamefont {L.~F.}\ \bibnamefont
  {Santos}},\ }\bibfield  {title} {\bibinfo {title} {{Integrability of a
  disordered Heisenberg spin-1/2 chain}},\ }\href
  {https://doi.org/10.1088/0305-4470/37/17/004} {\bibfield  {journal} {\bibinfo
      {journal} {J. Phys. A}\ }\textbf
  {\bibinfo {volume} {37}},\ \bibinfo {pages} {4723} (\bibinfo {year}
  {2004})}\BibitemShut {NoStop}%
\bibitem [{\citenamefont {Brenes}\ \emph {et~al.}(2020)\citenamefont {Brenes},
  \citenamefont {LeBlond}, \citenamefont {Goold},\ and\ \citenamefont
  {Rigol}}]{Brenes.2020}%
  \BibitemOpen
  \bibfield  {author} {\bibinfo {author} {\bibfnamefont {M.}~\bibnamefont
  {Brenes}}, \bibinfo {author} {\bibfnamefont {T.}~\bibnamefont {LeBlond}},
  \bibinfo {author} {\bibfnamefont {J.}~\bibnamefont {Goold}},\ and\ \bibinfo
  {author} {\bibfnamefont {M.}~\bibnamefont {Rigol}},\ }\bibfield  {title}
  {\bibinfo {title} {{Eigenstate Thermalization in a Locally Perturbed
  Integrable System}},\ }\href {https://doi.org/10.1103/physrevlett.125.070605}
  {\bibfield  {journal} {\bibinfo  {journal} {Phys. Rev. Lett.}\
  }\textbf {\bibinfo {volume} {125}},\ \bibinfo {pages} {070605} (\bibinfo
  {year} {2020})}\BibitemShut {NoStop}%
\bibitem [{\citenamefont {Santos}\ \emph {et~al.}(2020)\citenamefont {Santos},
  \citenamefont {Pérez-Bernal},\ and\ \citenamefont
  {Torres-Herrera}}]{Santos.2020}%
  \BibitemOpen
  \bibfield  {author} {\bibinfo {author} {\bibfnamefont {L.~F.}\ \bibnamefont
  {Santos}}, \bibinfo {author} {\bibfnamefont {F.}~\bibnamefont
  {Pérez-Bernal}},\ and\ \bibinfo {author} {\bibfnamefont {E.~J.}\
  \bibnamefont {Torres-Herrera}},\ }\bibfield  {title} {\bibinfo {title}
  {{Speck of chaos}},\ }\href
  {https://doi.org/10.1103/physrevresearch.2.043034} {\bibfield  {journal}
  {\bibinfo  {journal} {Phys. Rev. Res.}\ }\textbf {\bibinfo {volume}
  {2}},\ \bibinfo {pages} {043034} (\bibinfo {year} {2020})}\BibitemShut
  {NoStop}%
\end{thebibliography}%
\end{document}